\documentclass[fleqn,usenatbib,useAMS]{mnras}

\usepackage{graphicx}	
\usepackage{amsmath}	
\usepackage{amssymb}	
\usepackage{multicol}        
\usepackage{bm}		
\usepackage{pdflscape}	
\usepackage{subfigure} 

\usepackage[T1]{fontenc}
\usepackage{ae,aecompl}

\usepackage{newtxtext,newtxmath}

\title[MNRAS \LaTeX\ guide for authors]{\textit{An Alternative Formation Scenario for Uranium-rich Giants: Engulfing a Earth-like Planet}}
\author[Dian Xie]{Dian Xie,$^{1,}$ $^{2}$%
	Chunhua Zhu*,$^{1}$%
	Sufen Guo,$^{1}$%
	Helei Liu,$^{1}$%
	Guoliang L\"{u}*$^{2,}$ $^{1}$
\thanks{Contact e-mail: \href{mailto:mn@ras.ac.uk}{chunhuazhu@sina.cn}, {guolianglv@xao.ac.cn}}%
\thanks{Present address: School of Physical Science and Technology,Xinjiang University, Urumqi, 830017, China}
\\
$^{1}$School of Physical Science and Technology,Xinjiang University, Urumqi, 830017, China\\
$^{2}$Xinjiang Astronomical Observatory, Chinese Academy of Sciences, 150 Science 1-Street, Urumqi, Xinjiang 830011, China}
\date{Accepted 2023 July 10. Received 2023 July 6; in original form 2023 April 16}

\pubyear{2023}

\begin{document}
\label{firstpage}
\pagerange{\pageref{firstpage}--\pageref{lastpage}}
\maketitle

\begin{abstract}
The actinides, such as the uranium (U) element, are typically synthesized through the rapid neutron-capture process (r-process), which can occur in core-collapse supernovae or double neutron star mergers. There exist nine r-process giant stars exhibiting conspicuous U abundances, commonly referred to as U-rich giants. However, the origins of these U-rich giants remain ambiguous. We propose an alternative formation scenario for these U-rich giants whereby a red giant (RG) engulfs an Earth-like planet. To approximate the process of a RG engulfing an Earth-like planet, we employ an accretion model wherein the RG assimilates materials from said planet. Our findings demonstrate that this engulfment event can considerably enhance the presence of heavy elements originating from Earth-like planets on the surfaces of very metal-poor stars ($Z = 0.00001$), while its impact on solar-metallicity stars is comparatively modest. Importantly, the structural and evolutionary properties of both very metal-poor and solar-metallicity stars remain largely unaffected. Notably, our engulfment model effectively accounts for the observed U abundances in known U-rich giants. Furthermore, the evolutionary trajectories of U abundances on the surfaces of RGs subsequent to the engulfment of Earth-like planets encompass all known U-rich giants. Therefore, it is plausible that U-rich giants are formed when a RG engulfs an Earth-like planet.
\end{abstract}




\begingroup
\let\clearpage\relax
\endgroup
\newpage

\section{Introduction}

Recently, according to the data from \citet{Abohalima2018} in "JINAbase", there are nine giants that have been identified with clearly detectable U. These stars are known as U-rich giants and include: CS 31082-001 \citep{Hill2002}, BD+173248 \citep{Cowan2002}, CS 22892-052 \citep{Honda2004}, CS30306-132 \citep{Honda2004}, HD 115444 \citep{Honda2004}, HD186478 \citep{Honda2004}, HD6268 \citep{Honda2004}, HE 1523-0901 \citep{Frebel2007}, CS 29497-004 \citep{Hill2017}. These U-rich giants are very metal-poor (VMP) stars, that is, [Fe/H] <= -2. In this work, the notation used to represent elemental abundances in spectroscopy follows the standard notation as described by \citet{Helfer1959}.For elements X and Y, the notation is as follows:
\begin{equation}
	\log \varepsilon(\mathrm{X}) \equiv \log _{10}\left(\mathrm{~N}_{\mathrm{X}} / \mathrm{N}_{\mathrm{H}}\right)+12.0 \text, 
\end{equation}
where $\mathrm{~N}_{\mathrm{X}}$ and $\mathrm{N}_{\mathrm{H}}$ represent the number densities of element X and hydrogen, respectively.
\begin{equation}
	[\mathrm{X} / \mathrm{Y}] \equiv \log _{10}\left(\mathrm{~N}_{\mathrm{X}} / \mathrm{N}_{\mathrm{Y}}\right)_*-\log _{10}\left(\mathrm{~N}_{\mathrm{X}} / \mathrm{N}_{\mathrm{Y}}\right)_{\odot},
\end{equation}
where $\mathrm{~N}_{\mathrm{X}}$ and $\mathrm{~N}_{\mathrm{Y}}$ represent the number densities of elements X and Y, respectively.

U belongs to actinides which are believed to have originated predominantly from explosive r-process nucleosynthesis. The r-process is considered a significant mechanism for the production of elements heavier than iron(Fe) and is the only known process capable of synthesizing actinides. Depending on the degree of r-process enrichment, they can be classified into different categories: r-I stars: These stars have 0.3 <= [Eu/Fe] <= 1 and [Ba/Eu] < 0. They are believed to form in slightly larger dwarf galaxies, such as Tucana III \citep{Hanse2017}. r-II stars: These stars have [Eu/Fe] > 1 and [Ba/Eu] < 0. They are found in ultra-faint dwarf galaxies (UFD) like Reticulum II \citep{Beers2005, Ji2016, Roederer2016}. According to the data from \citet{Abohalima2018} in "JINAbase", there have been approximately 91 r-I stars and 32 r-II stars identified. The nine known U-rich giant belongs to r-I or r-II stars.

However, in the field of astrophysics, there are two main candidates that can produce actinides: core-collapse supernovae (CCSNe) and neutron star mergers (NSMs). Obviously, the U-rich giants cannot produce U by themselves. Therefore, their origin is still debated. \citet{Choplin2022} demonstrated that actinides can also be synthesized in low- metallicity, low-mass AGB stars through the i-process (the intermediate neutron capture process). However, their model result is strongly affected by the remaining uncertainties.  

It is widely acknowledged that planets exist in nearly all stellar systems, including our own solar system \citep{Mayor2011,Melendez2017,Zhu2018}. With the evolution of host star, it begins to expand. The host star can engulf its planets, and undergo a physical process like as common-envelope evolution \citep[e. g., ][]{Nordhaus2013}. This process is referred to as planetary engulfment. A number of literatures have investigated the impact of this process on their host stars. \cite{Livio1984} suggested that the substellar companions around stellar remnants can produced via planetary engulfment \citep{Bear2021,Yarza2022}. \cite{Soker1998} considered that the planetary engulfment can enhance the rotation of host star\cite[e. g.,][]{Siess1999, Privitera2016}. \cite{Sandquist1998} found that lithium enrichment on the surface of giant star can be explained via planetary engulfment\citep[e. g.,][]{Soares-Furtado2021}.  \citet{Ramirez2015} and \citet{Melendez2017} conducted research on a main sequence star that undergoes planet engulfment. They observed that a small convective region within the host star leads to an enrichment of heavy elements on its surface. Not only that the ingestion of planets can be deduced through the augmentation of refractory elements in the photosphere of the host star subsequent to the accretion of rocky planetary material. These enhancements of refractory substances are influenced by internal mixing mechanisms within stellar structures, specifically thermohaline mixing caused by an inverse gradient of mean molecular weight between the convective envelope and radiative core\citep[e. g.,][]{Melendez2009,Behmard2023}. Therefore, U-rich giants may be produced via the planetary engulfment.

In this paper, our primary emphasis lies in the investigation of host stars that engulf Earth-like planets during their red giant phase. We delve into the likelihood of these giants transforming into U-rich giants. Section 2 encompasses our comprehensive models regarding the process of a star engulfing a rocky planet. Section 3 entails a detailed analysis of both the Fe and U abundances subsequent to planetary engulfment. Ultimately, our conclusions are encapsulated within Section 4.

\section{Red Giants Engulfing Rocky Plants}
To investigate the process of a red giant engulfing a rocky planet, it is necessary to simulate both the stellar structure and evolution, as well as the interaction between the red giant and its planet. For this purpose, we employ the open-source evolutionary stellar code Modules for Experiments in Stellar Astrophysics (MESA; \citet{Paxton2011, Paxton2013, Paxton2015, Paxton2018, Paxton2019}, version 12115) to calculate the stellar evolution. In addition, we use an accretion model to simulate the engulfment of the red giant and its planet.

\subsection{Input Parameters for Stellar Evolution}
\begin{figure}       
	\quad
	\subfigure[]{
			\hspace{-0.3cm}\includegraphics[width=\columnwidth]{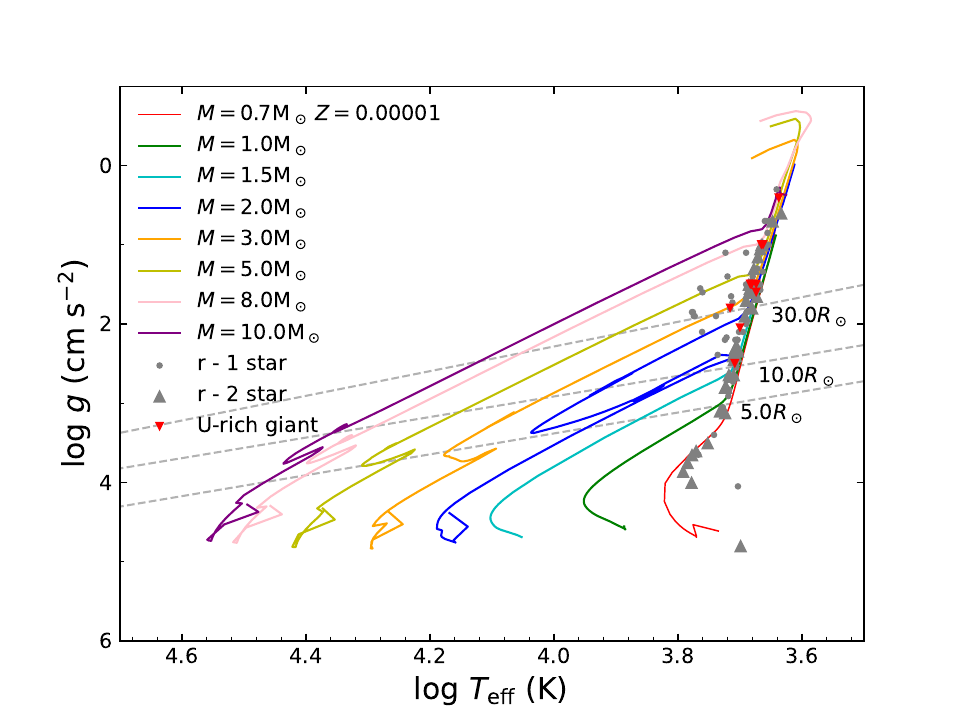}
	}
	\quad
	\subfigure[]{
		\includegraphics[width=\columnwidth]{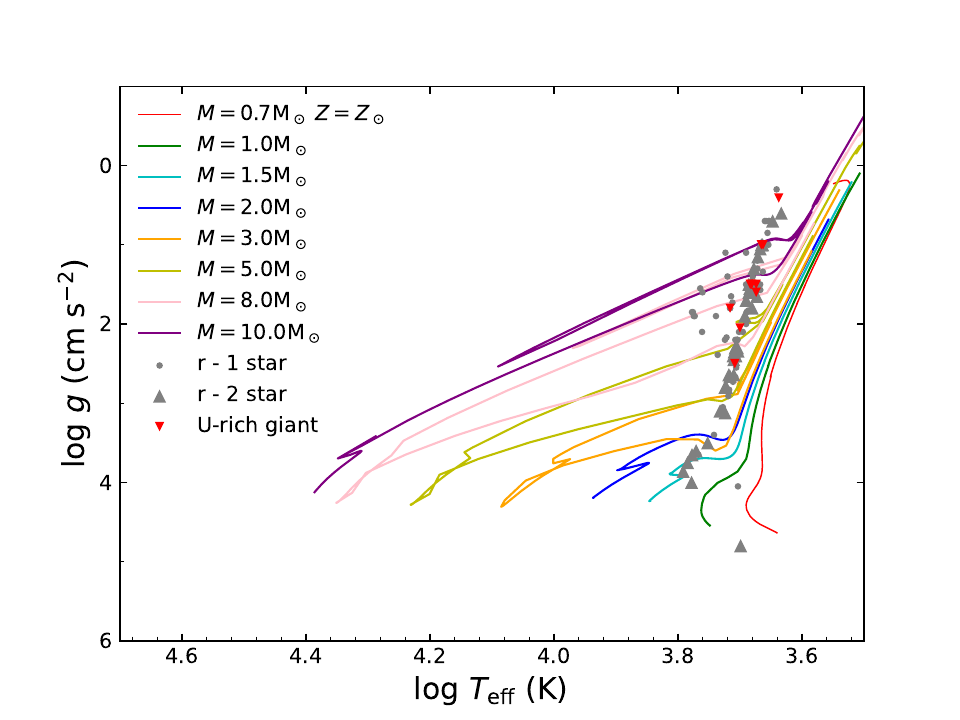}
	}
	\caption{Evolutionary tracks of the donors in the $T_{eff}$ -- log $g$ plane from the zero-age main sequence to the main sequence by 0.7 M$_\odot$, 1.0 M$_\odot$, 1.5 M$_\odot$, 2.0 M$_\odot$, 3.0 M$_\odot$, 5.0 M$_\odot$, 8.0 M$_\odot$, 10.0 M$_\odot$, respectively. The gray dots are r-I stars, the gray triangles are r-II stars, and the red inverted triangles are r-process stars with detected U. The dashed line lines represent equal radius lines of 5.0 R$_\odot$, 10.0 R$_\odot$ and 30.0 R$_\odot$, respectively.}
	\label{fig:hr}
\end{figure}

The stellar structure and evolution mainly depends on the stellar mass and metallicity. The observational sample we used is basically very metal-poor (VMP) stars or even extremely metal-poor (EMP) stars, so we use $Z$ = 0.00001 to perform stellar evolution calculations. Fig. \ref{fig:hr} shows that $Z$ = 0.00001 can cover U-rich giants very well. In order to discuss the effects of metallicity on the formation U-rich giants, we take $Z$ = $Z_\odot$ and 0.00001 in the different models. Besides,  Fig. \ref{fig:hr} shows that 1.0 M$_\odot$, 2.0 M$_\odot$ and 5.0 M$_\odot$ evolution tracks can basically cover our sample.

Usually, convection, overshoot, thermohaline mixing, element diffusion, and radiative levitation exert significant influence upon the structural dynamics and evolutionary trajectories of stars, particularly shaping the chemical abundance patterns discernible on their stellar surfaces. In the present paper, the Ledoux criterion is used for the convection. The mixing-length parameter $\alpha_{\mathrm{LMT}}=1.5$, the parameter of the semi-convection
$\alpha_{\mathrm{SEM}}=1.0$ \citep{Brott2011,Zhu2017,Cui2018,Lu2020}. The overshoot mixing diffusion coefficient that occurs near the convective boundary of a star is:
\begin{equation}
	D_{\mathrm{ov}}=D_{\mathrm{conv}, 0} \exp \left(-\frac{2 z}{f \lambda_{P, 0}}\right),
\end{equation}
where $D_{\mathrm{conv}, 0}$ is the diffusion coefficient near the Schwarzschild boundary, $\lambda_{P, 0}$ is height of the pressure scale in this position, ${z}$ is the distance in the radiation layer away from this position, and $f$ is a parameter which may have different values at the upper and lower convective boundaries for no-burning, H-burning, He-burning, and metal-burning convection zones \citep{Herwig2000}. For simplicity, $f=0.02$ in our models.

Thermohaline mixing occurs in the presence of inversions, where regions with an inverted average molecular weight are considered formally stable. The diffusion coefficient is determined through linear stability analysis by \citet{Ulrich1972} and \citet{Kippenhahn1980}. This type of mixing is particularly significant in cases of planetary engulfment, where heavy planetary material is deposited near the star's surface. In this study, we applied the method developed by \citet{Brown2013}, which is based on \citet{Kippenhahn1980} and provides a more comprehensive and precise approach to investigate thermohaline mixing.

Due to the large number of calculations required for diffusion calculations for each species, MESA groups species into different categories for diffusion calculations \citet{Paxton2018}.  Hydrogen and deuterium would be placed in `$^{1}$H' and carbon, nitrogen and oxygen would be placed in `$^{16}$O', and anything heavier in `$^{56}$Fe'. Apparently U is also treated as `$^{56}$Fe' for element diffusion. We turned it on.

Meanwhile, we have also accounted for radiative buoyancy in our model \citep{Hu2011}. Radiative levitation is a phenomenon in stellar atmospheres where the radiation pressure from intense radiation fields can push elements upwards, affecting their distribution and abundance. By introducing this extra force term into the existing models, \citet{Hu2011} aim to more accurately account for the effects of radiative levitation on the dynamics and composition of stellar atmospheres. They incorporate an additional force component attributed to radiative levitation.

\subsection{Engulfing Planet Model } \label{Accretion of Mass Change}
The host star engulfing its planet has been investigated by many literatures \citep{Nelemans1998,Metzger2012,Qureshi2018, Salas2019}. They suggested that the planet should be dissolved and its matter should be added to the host star by a combination of ram pressure and tidal forces near the base of the convective envelope.

\citet{Nelemans1998} investigated that a solar-like star engulfed its planet, and they considered that the planet dissolve at the position where the local sound speed $ c_{\rm s}$ in the stellar envelope equals the escape velocity $v_{\mathrm{esc}}$ at the planet surface, that is:
\begin{equation}
	c_{\mathrm{s}}^2 \approx v_{\mathrm{esc}}^2 \Longleftrightarrow \gamma \frac{k_{\mathrm{B}} T}{\mu m_{\mathrm{u}}} \approx \frac{2 G m_{\mathrm{p}}}{\alpha r_{\mathrm{p}}}
	\label{FDF/FD},
\end{equation}
where $m_{\rm p}$ and $r_{\rm p}$ are the mass and radius of the planet, respectively. Here, parameters, $\alpha$ = 1 \citep{Nelemans1998}. For an Earth-like planet ($m_{\rm p}=5.9\times10^{21}$ g and $r_{\rm p}=6.3\times10^{8}$ cm), $\log c_{\rm s}\sim$ 6.1 cm s$^{-1}$.

\begin{figure}    
	\quad
	\subfigure[]{
			\hspace{-0.3cm}\includegraphics[width=\columnwidth]{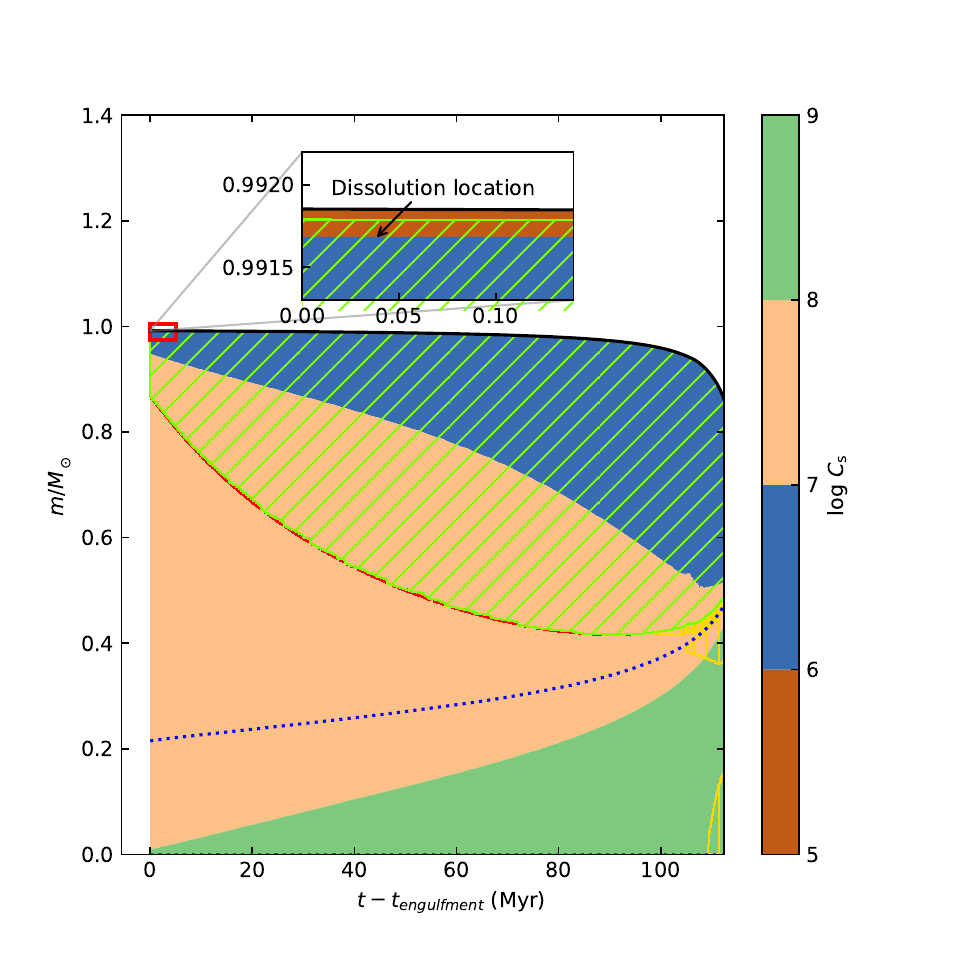}
	}
	\quad
	\subfigure[]{
		\includegraphics[width=\columnwidth]{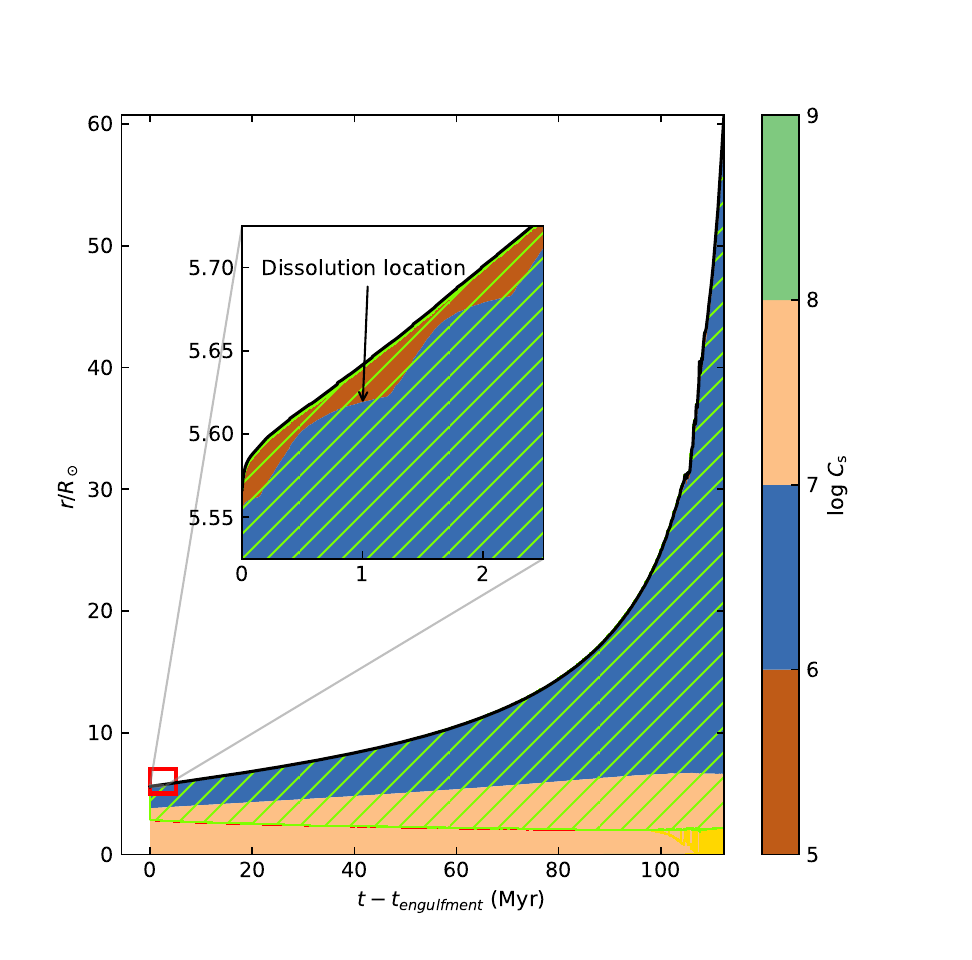}
	}
	\caption{Local sound velocity ($\log c_{\rm s}$) diagram along the stellar mass (panel (a)) and radius (panel (b)) vs. stellar evolution for 1 M$_\odot$ model with $Z=0.00001$. Color zones show the value ranges of different $c_{\rm s}$. The green shading zone represents the convection zone, the blue dotted lines give the helium core zone. The positions where the Earth-like planet is dissolved (that is, $\log c_{\rm s}=6.1$), is located at the intersection between the blue and brown zone.  Their details are given sub-panels.}
	\label{csound}   
\end{figure}

The $c_{\rm s}$ in a stellar envelope is dependent on the structure of the star. Fig. \ref{csound} shows $c_{\rm s}$ profiles of red giant. Obviously, based on Eq. \ref{FDF/FD}, the planet is dissolved at a mass thickness of about 10$^{-5}$ M$_\odot$ (or a depth of about 10$^{-2}$ R$_\odot$) under the stellar surface. It means that the planet dissolution only occurs at zone very closed to the stellar surface, which is consistent with the results of \citet{Lau2022}. Therefore, in the present paper, we use an accretion model to approximate the planet dissolving in its host star.

The mass-accretion rate of the accretion model can be approximately estimated via the mass-dissolving rare of the planet. \citet{Church2020} investigated the ingestion of planets into the surface layers of the star, and calculated the critical condition for dissolution and the mass-dissolving rate of the planet. Based on the drag force ($F_{\mathrm{D}}$) and the gravitational binding energy of the planet surface ($\epsilon_{\text {bind }, \mathrm{p}}$), \citet{Church2020} suggested that the mass-dissolving rate of the planet could be given by

\begin{equation}
	\dot{M}_{\mathrm{p}}=\frac{C_{\mathrm{H}} F_{\mathrm{D} }v}{\epsilon_{\text {bind }, \mathrm{p}}+\mathcal{L}_{\mathrm{vap}}}
	\label{Mp}
\end{equation} where $F_{\mathrm{D}}=\frac{1}{2} C_{\mathrm{D}} \pi r_{\mathrm{p}}^2 \rho_{\star}(r) v^2$, $\epsilon_{\text {bind }, \mathrm{p}} \simeq \frac{G m_p}{r_p}$ and $\mathcal{L}_{\mathrm{vap}}$ is the latent heat of vaporization. Here, $v$ is the planet  velocity relative to the host star. The drag coecient $C_{\mathrm{D}}$ = 1, $C_{\mathrm{H}}$=0.01 \citep{Church2020}. The density of the stellar surface layer $\rho_{\star}$ is obtained by MESA. Considered that the planet is the Earth-like planet, $\mathcal{L}_{\mathrm{vap}}$ is the latent for Fe, that is, $\mathcal{L}_{\mathrm{vap}}$$\simeq$ 6 kJ g$^{-1}$.

According to Eq. \ref{Mp}, we can estimate that $\dot{M}_{\rm p}=5\times10^{-8}$ M$_\odot$ yr$^{-1}$ when the Earth is engulfed in the stellar envelope. It indicates that the dissolving process of the Earth lasts about $10^3$ yr. When the materials from the Earth is dissolved into the stellar envelope, the heavy elements are mixed whole convective region. For a region where the diffusion velocity ${v_i}$ of an element $i$ is relatively constant, the mixing timescale can be estimated by
\begin{equation}
	\tau \approx \frac{l}{v_i}=\frac{l p}{\rho g D}=\frac{l H_p}{D},\label{6}
\end{equation}where $l$ is the width of the convective, and it is about $3.0$ $R_\odot \sim 50.0$ $R_\odot$ for a red giant showed in Fig. \ref{csound}. ${H_p}$ and $D$ are the pressure scale height and the mixing coefficient, and they are $10^{-1.2}R_\odot$ and $10^{6.5}$cm$^2$s$^{-1}$ for a red giant, respectively. Therefore, the mixing time scale $\tau \approx 10^6$$\sim$$10^7$ yr. Compare with the timescale of dissolving planet ($\sim10^3$ yr), the mixing timescale is very long. It indicates that the U-rich giants observed mainly are in mixing phase after the planet is dissolved.

Simultaneously, the chemical composition of accreted materials is very important. In this work, the planet is the Earth-like planet, and we focus on the formation of U-rich giant. Therefore, the chemical abundances of main heavy elements plus Th and U are similar to those in the Earth, which are listed in Table \ref{tab:mass fractions}.

\begin{table}
	\centering
    \caption{Mass fraction of main elements from \citet{McDonough1995}, Th and U for an Earth-like planet.}
	\begin{tabular}{ll}
		\hline
		\hline
		Element & Mass Fraction $(\%)$ \\
		\hline
		Fe & 37.5 \\
		O  & 30.1     \\
		Si & 15.1     \\
		Mg & 13.9      \\
		Ni & 1.8      \\
		Ca & 1.5      \\
		Th & $6\times10^{-7}$    \\
		U  & $1.4\times10^{-6}$    \\
		
		\hline
	\end{tabular}
	\label{tab:mass fractions}
\end{table}

\section{Formation of Uranium-rich giants via engulfing a rocky planet}
The evolutionary phase of the host star may affect the planet dissolution. For a red giant, its radius becomes larger with its evolution. The dashed lines in Fig. \ref{fig:hr} (a) are the iso-radius lines for $r=$ 5.0, 10.0 and 30.0 R$_\odot$. Assuming that the red giant engulfing its planet occurs at different radii ($r=$ 5.0, 10.0 and 30.0 R$_\odot$, respectively), we have carried out some tests for effects of the giant radius on the formation of U-rich giant, and find that the effects are very weak. Therefore, in all simulation, we assume that the engulfment begins when the red giant has a radius of 5.0 R$_\odot$ .

\subsection{Evolution and Effects of Fe Element}
As shown by Table \ref{tab:mass fractions}, Fe is the most abundant element in the Earth-like planet. It is well known, Fe element plays a significant role in the stellar structure and evolution. It is necessary to discuss its evolution and effects on the host star after an Earth-like planet is engulfed.

\begin{figure}
	\includegraphics[width=0.9\columnwidth]{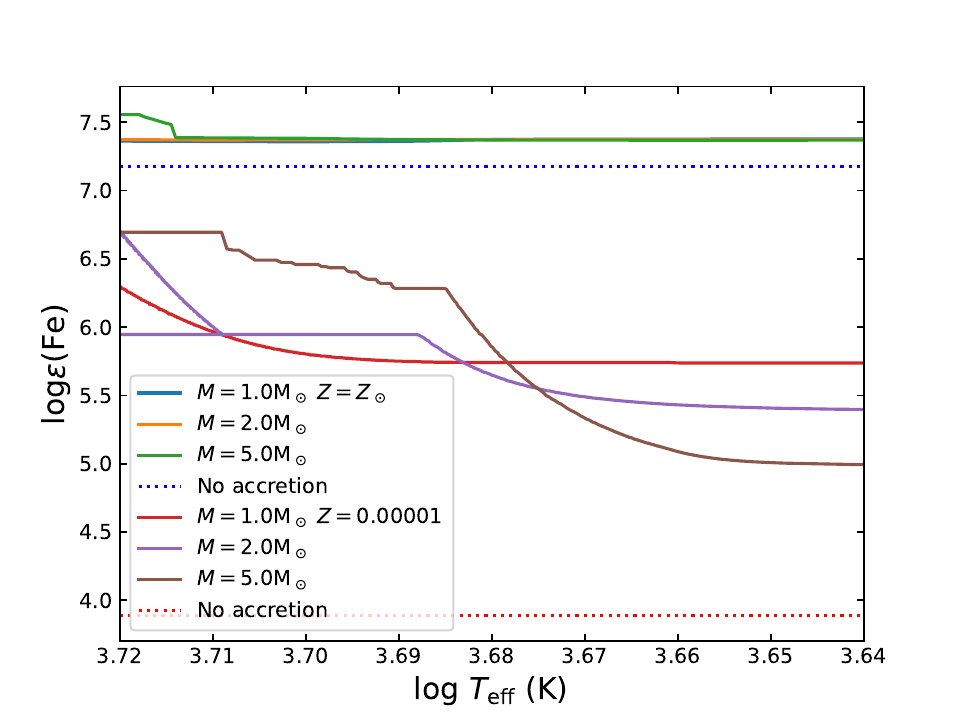}
	\caption{Evolution of Fe abundance with the ective temperature for the engulfing models involving 1.0 M$\oplus$ planet. The solid lines in different colors represent changes in Fe abundance after accretion of models with different metallicity and different initial masses. The blue and red dotted lines represent the Fe abundance in a model with non-accretion, $Z$ = $Z_\odot$ and $Z$ = 0.00001, respectively.}
	\label{fig:fe}
\end{figure}

Fig. \ref{fig:fe} shows the evolution of Fe abundance with the effective temperature for the engulfing models involving 1.0 M$\oplus$ planet. Obviously, after engulfing an Earth-like planet, the Fe abundance on the surface of host star enhances. Especially, in the models with $Z=0.00001$, it increases by hundreds of times because these host stars have very low Fe abundance before the engulfment. With the evolution of the host star, Fe elements accreted are gradually mixed whole envelope. Therefore, Fe abundance reduces bit by bit. Compared with  the Fe abundance on the stellar surface of the models with the solar metallicity,  it for the models with $Z=0.00001$ has a significant decrease due to very low initial Fe abundance within stellar envelope.

\begin{figure}
	\includegraphics[width=0.9\columnwidth]{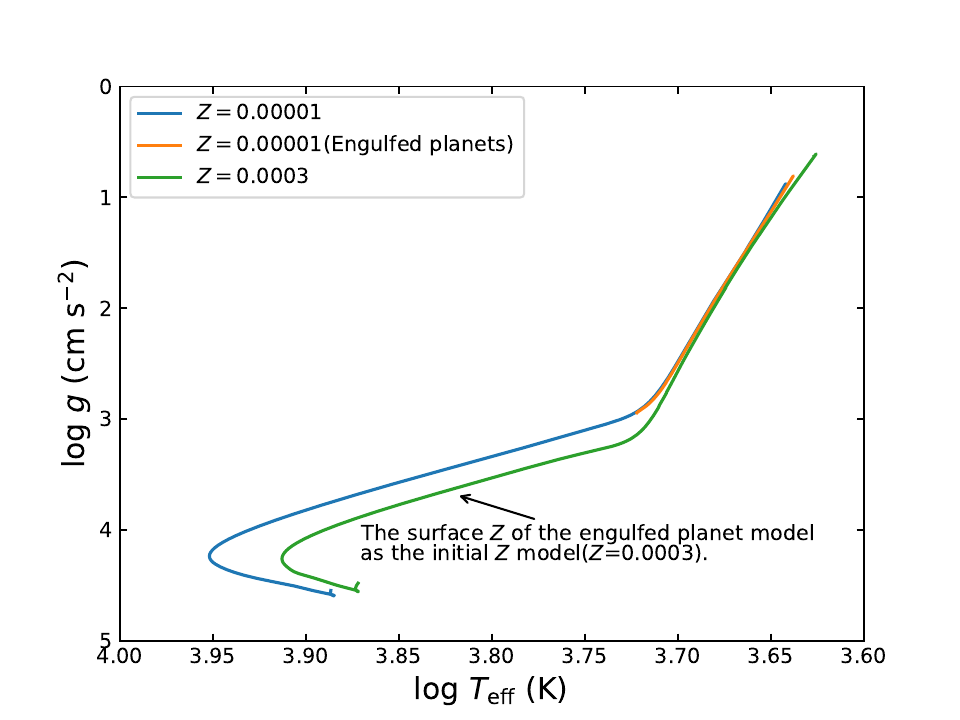}
	\caption{The blue solid line and the orange solid lines are the evolution curves for the models with 1.0 M$_\odot$ star with $Z=0.00001$ after not engulfing or engulfing an Earth-like planet, respectively. The green line shows the evolution of the model with $Z=0.0003$ which approximately equals the metallicity of 1.0 M$_\odot$ with $Z=0.00001$ engulfing an Earth-like planet.}
	\label{fig:acc}
\end{figure}

As the above discussions, the engulfing an Earth-like planet significantly changes Fe abundance in the envelope of VMP star. However, as the blue and red lines in Fig. \ref{fig:acc} which gives the evolutionary tracks of a 1.0 M$_\odot$ star with $Z=0.00001$ after not engulfing or engulfing an Earth-like planet show, it has no effects on the stellar evolution.

\subsection{U-rich giant formation via the engulfment of an Earth-like planet}
Although the U abundance in an Earth-like planet is very low ($\sim 10^{-6}$, See Table \ref{tab:mass fractions}), it still is very higher than those in the known U-rich RGs. Therefore, when a host star engulfs an Earth-like planet, the U abundance on its surface may be enhanced, and it may become the U-rich RG on observations. In general, the U abundance in the engulfment progress depends on the mass of the Earth-like planet. Typically, their masses are several the Earth mass \citep{Melendez2017,Liu2018,han2023}. \citet{Kishalay2023} reported the engulfment of a planetary body with a mass approximately ten times that of Jupiter by a solar-like star in the case of ZTF SLRN-2020. In order to discuss the effects of the planet mass on the formation of the U-rich RGs, we take 0.5 M$\oplus$, 1.0 M$\oplus$, 2.0 M$\oplus$, 5.0 M$\oplus$, 8.0 M$\oplus$ and 10.0 M$\oplus$ as the Earth-like planet’s mass in the different simulations.

\begin{figure}
	\includegraphics[width=0.9\columnwidth]{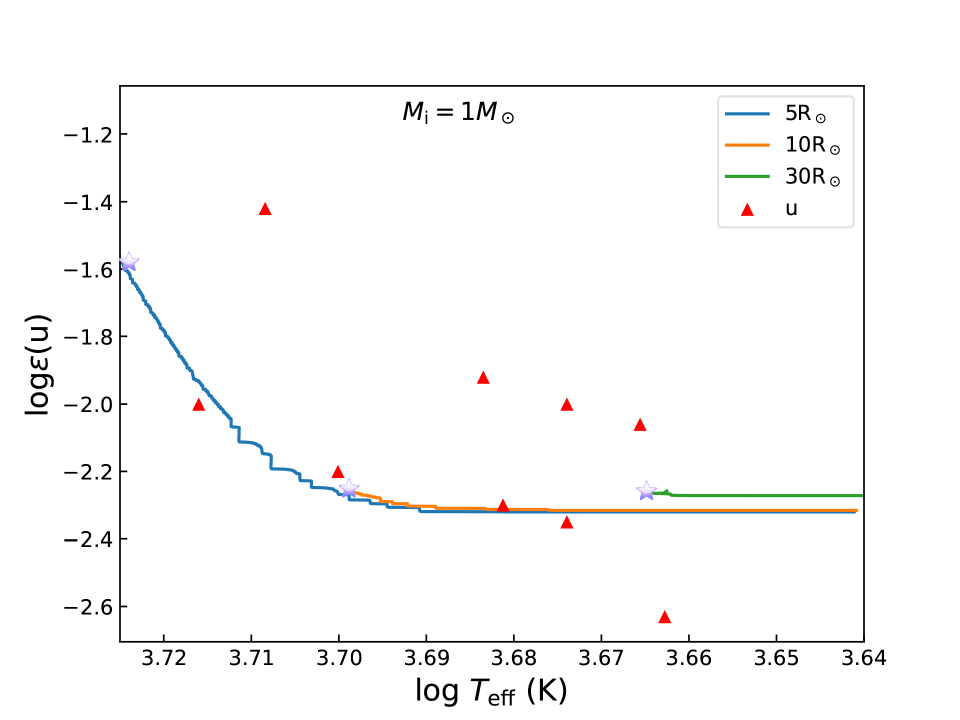}
	\caption{The evolution of U abundance on the RG surface with the effective temperature after the engulfment. The purple pentagon symbols represent the initiation of accretion at different solar radii. The blue, orange, and green lines correspond to the models with initial masses of 1M$_\odot$ from 5R$_\odot$, 10R$_\odot$, and 30R$_\odot$ involving 1.0 M$\oplus$ Earth-like planet engulfment, respectively.}
	\label{fig:radii}
\end{figure}

Simultaneously, the structure of the host star may also affect the U abundance during the engulfment process. For a RG, its radius becomes larger with its evolution. The dashed lines in Fig. \ref{fig:hr} (a) are the iso-radius lines for $r$ = 5.0, 10.0 and 30.0 R$_\odot$. Assuming that the red giant engulfing its planet occurs at different radii ($r$ = 5.0, 10.0 and 30.0 R$_\odot$, respectively), we have carried out some tests for effects of the giant radius on the formation of U-rich giant, and found that the effects are very weak. Fig. \ref{fig:radii} shows the evolution of U abundance on the stellar surface when the planetary engulfment occurs at different stellar radii, that is, R=5.0, 10.0 and 30.0 R$_\odot$. Obviously, U abundance is higher at the case of R= 5 R$_\odot$. The main reason is that the convective envelope at this time is the smallest. However, with the stellar evolution, U abundances at three cases become very closed. Simultaneously, based on observations, some U-rich giants can not be covered if the planetary engulfment occurs too late. Explicitly stating that RGs with smaller radii/thinner convective envelopes exhibit initially higher U-enhancements (as expected), but this effect levels out as stellar evolution proceeds. Therefore, in all simulations, we assume that the engulfment begins when the red giant has a radius of 5.0 R$_\odot$.

\begin{figure}
    \includegraphics[width=0.8\linewidth]{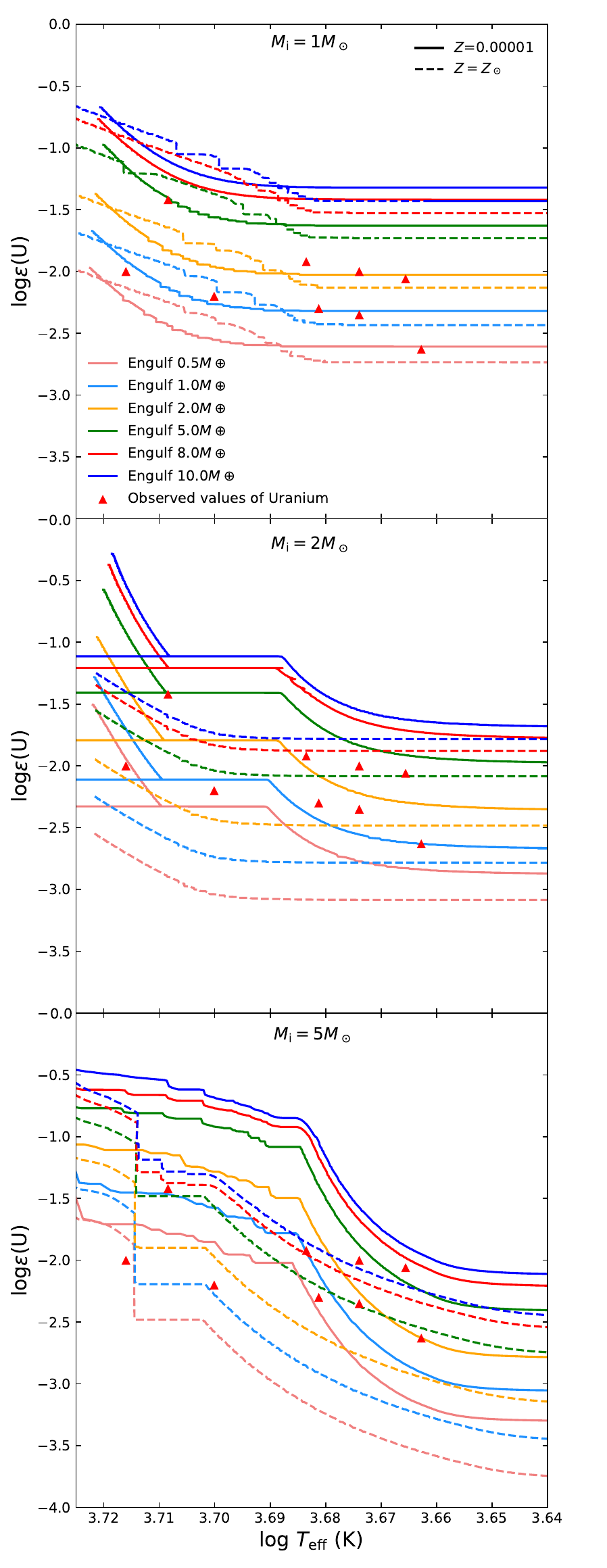}
	\caption{The evolution of U abundance on the RG surface with the effective temperature after the engulfment. The red, blue, yellow and green lines show the evolutions for the models involving 0.5 M$\oplus$, 1.0 M$\oplus$, 2.0 M$\oplus$, 5.0 M$\oplus$, 8.0 M$\oplus$ and 10.0 M$\oplus$ Earth-like planet engulfment, respectively. The solid and dashed lines represent the models with $Z=0.00001$ and $Z=Z_\odot$, respectively. The left, center and right panels correspond to the models of the host stars with an initial mass of 1.0 M$_\odot$, 2.0 M$_\odot$ and 5.0 M$_\odot$, respectively.}
    \label{u} 
\end{figure}

Fig. \ref{u} shows the evolution of U abundance on the RG surface with the effective temperature after the engulfment. Obviously, our simulating results can cover all known U-rich RGs. The larger the Earth-like planet’s masses are, the higher the U abundances on the surface of host stars are. The models engulfing a 5.0 M$\oplus$  Earth-like planets can explain the U abundances of CS30306-132 with log $\varepsilon$ (U)$\sim$ -1.42. There are some super-Earth planet discovered \citep{han2023}. Therefore, a RG engulfing an Earth-like planet can become an U-rich giant.

\begin{figure}
	\includegraphics[width=0.8\linewidth]{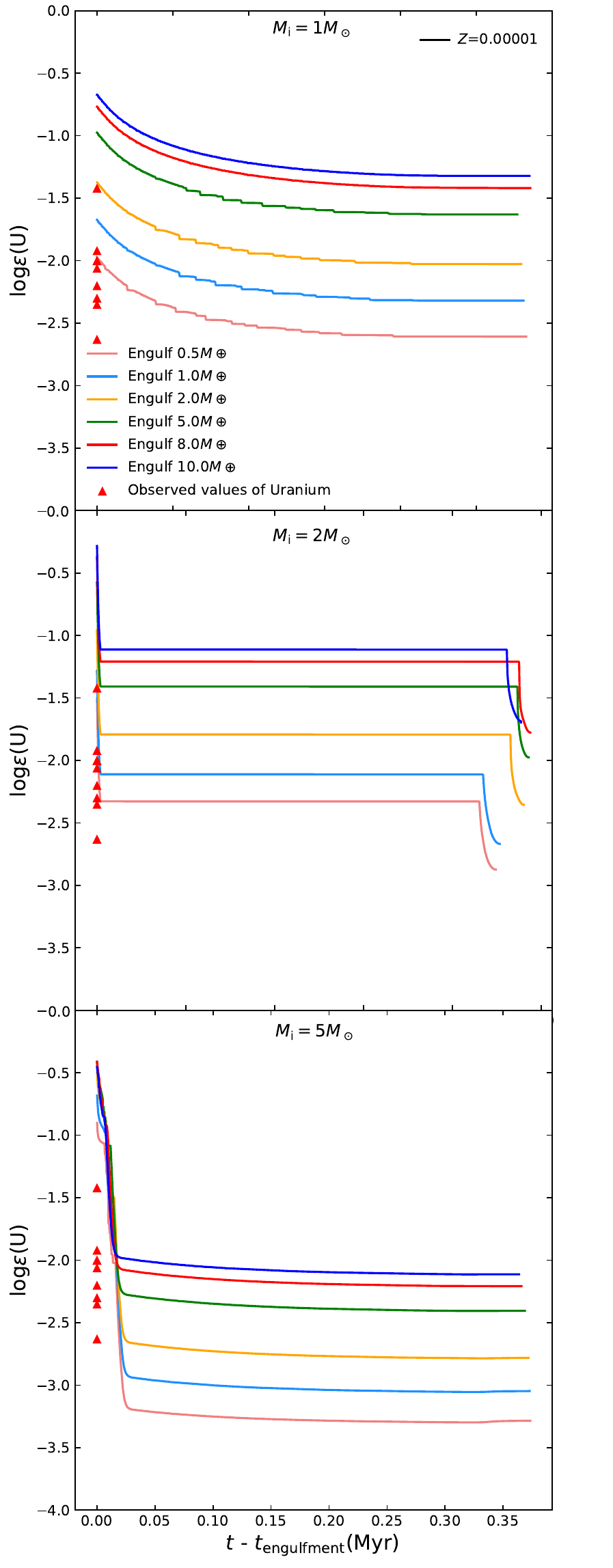}
	\caption{ Evolution of U abundance on the stellar surface after planetary engulfment. The red, blue, yellow, green and ultramarine blue lines show the evolutions for the models involving 0.5 M$\oplus$, 1.0 M$\oplus$, 2.0 M$\oplus$, 5.0 M$\oplus$, 8.0 M$\oplus$ and 10.0 M$\oplus$ Earth-like planet engulfment, respectively. The red triangles represent observations of U-rich giants, with their positions in the diagram solely indicative of U abundance.}
	\label{age} 
\end{figure}

Fig. \ref{age} shows the evolution of U abundance on the stellar surface after planetary engulfment. In our model, its evolution main depends on the timescale of the mixing and the mass of convective zone. After the planetary engulfment, U element from the Earth is dissolved into the stellar convective envelope. The timescale can be estimated by Eq. \ref{6}, and it approximately equals $10^6\sim$$10^7$ yr, which is consistent with Fig. \ref{age}. After the U element is homogeneously mixed within the convective envelope, U abundance keeps constant.

\begin{figure*}
	\quad
	\subfigure[]{
			\hspace{-0.3cm}\includegraphics[width=\columnwidth]{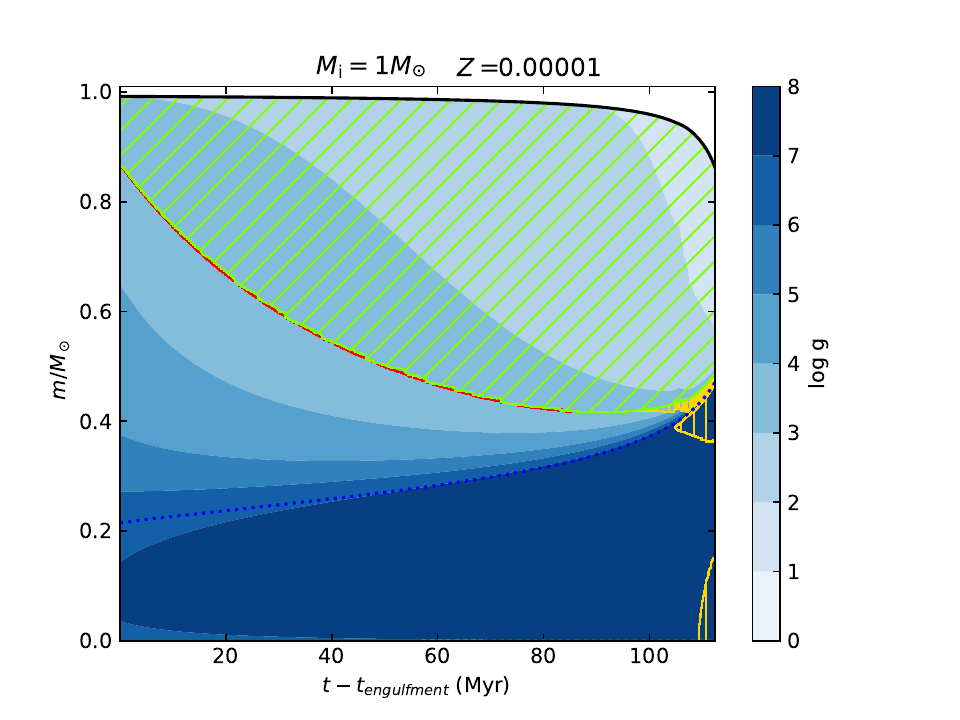}
	}
	\quad
	\subfigure[]{
		\includegraphics[width=\columnwidth]{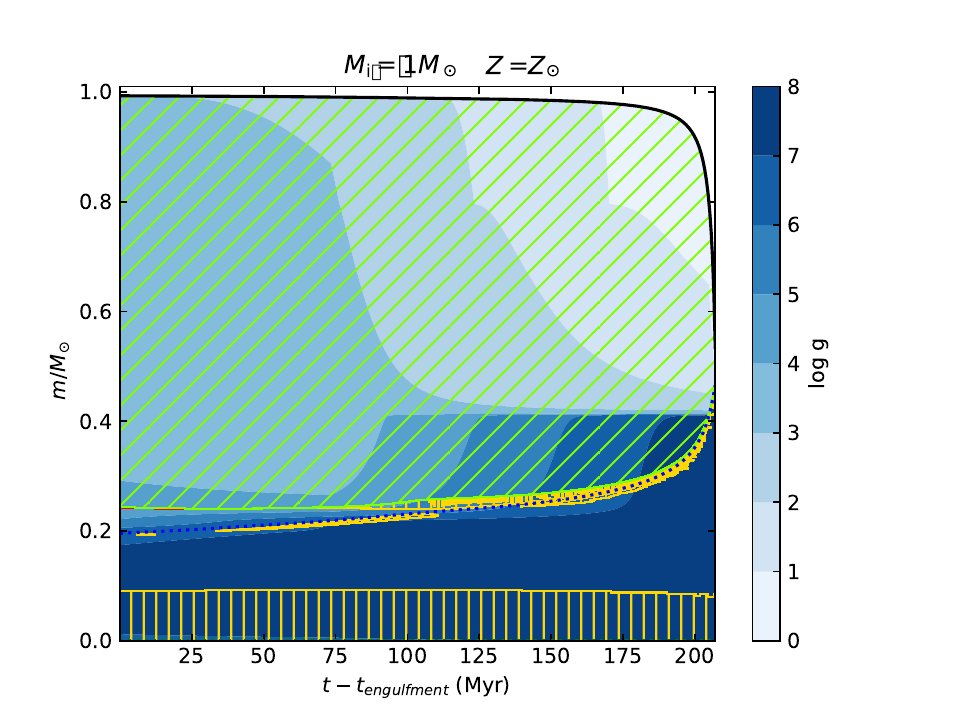}
	}
	\caption{Convection histories of the host stars with an initial mass of 1 M$_\odot$ after engulfment. The left and right panels are for the models with $Z=0.00001$ and $Z=Z_\odot$, respectively. The green shaded area is the convection mixing zone. The yellow shaded area is the  thermohaline mixing zone. The blue dashed line is helium nuclei. Color bar represents the value of gravity acceleration.}
	\label{kippa}      
\end{figure*}

As shown in Fig. \ref{u}, in our simulations, the U abundance in the models with $Z=Z_\odot$ is lower than that with $Z=0.00001$. The main reason is that, for a given stellar radius, the VMP RG has a convective zone smaller than that with solar metallicity. In our model, Fig. \ref{kippa} shows the convection histories of the 1 M$_\odot$ initial mass host stars with $Z$=0.00001 and $Z=Z_\odot$ after engulfment, respectively. The yellow area represents the thermohaline mixing zone. Obviously, the thermohaline mixing zone consists of two distinct regions, albeit occupying a relatively small proportion. One portion is located at the base of the upper convective zone, while another exists within the core region. Heavy elements are transported by convection to the thermohaline mixing zone at the base of the upper convective zone. However, there is no direct connection, making it challenging for elements to reach the core. Consequently, U abundance change caused by thermohaline mixing may not be significant in our simulations.

On the observations, all U-rich giants are VMP stars. However, based on our simulations, after engulfing an Earth-like planet, both the VMP or the solar-metallicity stars can become the U-rich giants. The possible reasons maybe the observational bias. Usually, the emission lines of solar-metallicity stars are so abundant that it is extremely difficult to detect U emission lines even though these stars engulf an Earth-like planet (See Fig. \ref{fig:fe}, that is, the engulfment has a very weak effect on the element abundances of the solar-metallicity stars.). The VMP stars have much low metal abundances, and the heavy-element abundances (especially U ) from the Earth-like planets are greatly enhanced after the engulfment. Therefore, it may be possible to detect U emission lines. However, this result needs to be supported by the further observations. For example, we can find some relics for the U-rich giants engulfing a planet, or can detect the abundance distribution of the heavy elements from the Earth-like planets, which can be explained by the engulfment model.

In addition, one should note, according to our current stellar planet engulfment mode, U element primarily comes from the Earth-like planets. The chemical compositions of the planets depend on the environment in which the host star is born. However, the formation environment of the star is determined by the previous generation of stars. The Sun should be the second generation star at least. The processes such as  CCSNe or NSMs should have occurred in the previous environment of the solar system,  which leads to the presence of gold, U and other products of these energetic processes on Earth. However, if the previous generation of a star did not experience CCSNe or NSM, the entire stellar system would lack actinide elements.  U-rich giant would not form through planetary engulfment. Maybe, the origin of U-rich (or other heavy elements) giants offers a potential avenue for studying the previous generation of stellar systems scientifically. 

Unfortunately, although there are 5,445 exoplanets observed \citep{Akeson2013, Shabram2020}, their chemical (especially heavy elements) abundances hardly are measured. Our knowledge to the heavy elements of exoplanets is extremely scarce. Therefore, we can only simulate the entire engulfment process using the compositions of the Earth.  Our model fails to apply to exoplanets without U or with extremely low U abundances.

\section{Conclusions}
We employ MESA as a computational tool to simulate the formation scenario of uranium-enriched giants. In light of the fact that planet dissolution predominantly occurs in the vicinity of the stellar surface during engulfment, we adopt an accretion model wherein a RG assimilates materials from an Earth-like planet. This approximation effectively represents the scenario where an RG engulfs an Earth-like planet. Our findings reveal that such engulfment can substantially augment the abundances of heavy elements originating from Earth-like planets on the surfaces of VMP stars ($Z=0.00001$), while its impact on solar-metallicity stars is comparatively modest. The structural and evolutionary characteristics of both VMP and solar-metallicity stars remain largely unaffected. Notably, our engulfment model adequately accounts for the observed U abundances in known uranium-rich giants. The evolutionary trajectories of U abundances on the surfaces of RGs after engulfing Earth-like planets encompass the entire population of known uranium-rich giants. Hence, it is plausible for a red giant to be formed through the engulfment of an Earth-like planet. However, further observational evidence is crucial to substantiate this formation mechanism for uranium-rich giants. 

\section*{Acknowledgments}
This work received the generous support of
the National Natural Science Foundation of China, project Nos. 12163005, U2031204 and 11863005,
the science research grants from the China Manned Space Project with NO. CMS-CSST-2021-A10,
the Natural Science Foundation of Xinjiang No.2021D01C075 and No.2020D01D85.

\section*{DATA AVAILABILITY}
R-process stars data are publicly available from "JINAbase" (\url{https://jinabase.pythonanywhere.com/}). Exoplanet data from \url{https://exoplanetarchive.ipac.caltech.edu}. Evolutionary models were computed with the version 12115 of MESA. The required inlists in this study are available via reasonable request to the corresponding author.


\bibliographystyle{mnras}
\bibliography{xd} 


\bsp	
\label{lastpage}
\end{document}